# Influence of conjugated structure for tunable molecular plasmons in peropyrene and its derivatives

Haoran Liu[1, #], Nan Gao[1, #], Yurui Fang[1, *]

[1] *School of Physics, Dalian University of Technology, Dalian* 116024, *China*

\* *Corresponding authors:* yrfang@dlut.edu.cn *(Y.F.)*
\# *The author contributed equally.*

**Abstract:** Advances in research have sparked an increasing curiosity in understanding the plasmonic excitation properties of molecular-scale systems. Polycyclic aromatic hydrocarbons, as the fundamental building blocks of graphene, have been documented to possess plasmonic properties through experimental observations, making them prime candidates for investigation. By doping different elements, the conjugated structure of the molecule can be altered. In this study, the plasmonic excitation properties influenced by conjugated structures in peropyrene and its derivatives are investigated through first-principles calculations that combine the plasmonicity index, generalized plasmonicity index and transition contribution maps. For molecular plasmonic excitation, the conjugated structure can influence the oscillation modes of valence electrons, which is pivotal in yielding distinct field enhancement characteristics. Furthermore, charge doping can lead to a certain degree of alteration in the conjugated structures, and the doping of elements will result in varying degrees of such alteration, thereby initiating different trends in the evolution of plasmonic resonance. This further enhances the tunability of molecular plasmonic resonance. The results provide novel insights into the development and utilization of molecular plasmonic devices in practical applications.

# 1. Introduction

Plasmons can be regarded as a quantized charge density wave that exhibit collective oscillations of conduction electrons, providing an extremely effective mechanism for confining and manipulating light at subwavelength scales.[1,2] Noble metal nanoparticles, renowned for their exceptional plasmonic optical properties at the nanoscale, have been the subject of study for centuries.[3] With the advances of research, there is a growing fascination with the properties and potential applications of plasmons in molecular-scale systems, including plasmonic catalysis,[4-6] quantum plasmonic excitations[7-9] and strong plasmon-molecule coupling.[10-16] Due to size constraints, quantum effects become more pronounced, leading to a transition of the density of states (DOS) from a continuous state to a set of discrete energy levels.[17] In this context, the collective excitations of correlated electrons in numerous systems, termed molecular plasmons, further pave a route for understanding and applications the plasmonic phenomenon in nanoscale limit. Thus, numerous research groups have investigated the molecular plasmonic excitation features.[18-32] Distinct plasmonic excitation modes have been observed in various structures such as metallic and semiconductor nanoclusters,[18,23,24,33,34] metal atomic chains,[18,23,35,36] alkene chains[18-20,37] and polycyclic aromatic hydrocarbons (PAHs).[23,24,32,38], where PAHs have as stronger plasmonic excitation as alkene chains due to the conjugated structures.

PAHs, as the fundamental building blocks of graphene — a material with two-dimensional collective excitation properties and excellent electro-optical characteristics [39-42] — characterized by their minimal size, which enables them to break the limitations of terahertz and mid-infrared resonance frequencies, allowing for the manifestation of molecular plasmonic resonances within the visible light region. The molecular plasmonic resonance frequency and intensity of PAHs are extremely sensitive to charge states, a fact that has been substantiated through experimental evidence.[38,43] Peropyrene[44,45] and its derivatives,[46] as members of PAHs, serve as excellent models for investigating plasmonic resonances and their tunability. While the collective tuexcitations in PAHs pave the way for innovative design and utilization of molecular plasmonic devices, a comprehensive grasp of their electrical tunability is essential to amplify the practical utility of these compact atomic structures.

Time-dependent (TD) density functional theory (DFT),[47,48] including linear response TDDFT (LR-TDDFT) and real-time TDDFT (RT-TDDFT), stands as a formidable computational tool, offering a platform for investigating the excitation of electrons in molecules. In this paper, we employ TD-DFT computational method to systematically investigate the plasmonic excitation and optical properties of peropyrene and its two derivatives. Using plasmonicity index (PI),[26] generalized plasmonicity index (GPI)[24] and transition contribution map (TCM)[25] methods to distinct the plasmonic excitation types in peropyrene and its derivatives, we find that there coexists both plasmonic and single-particle excitations in the weak dipole excitation modes on contrary of reported results,[49] which uncovers complexity and diversity in the optical behavior of these molecules. The plasmonic resonance modes are influenced by the conjugated structure of the molecular systems, with the valence electrons participating in the plasmonic oscillations predominantly being π-conjugated electrons. Replacing the ends atoms by only nitrogen atoms does not change the conjugated structure, a slight decrease in molecular size occurs, which causes a modest blueshift of longitudinal plasmonic resonance peak. The further doping of oxygen atoms hinders the degree of π electrons delocalization, leading to a noticeable redshift in the resonance energy and a significant decrease in peak intensity. Additionally, the modification of the conjugated structure results in a change in the electron oscillation modes practicing in the plasmonic excitation, thereby giving rise to distinct field enhancement characteristics. This provides a theoretical foundation for realizing nano-antenna structures with specific functionalities. Finally, charge doping can influence the conjugated structure of the system, leading to the evolution of the molecular plasmonic resonance modes in different directions. These research findings not only enhance our understanding of plasmons properties in molecules, but also provide significant scientific evidence for future applications in fields such as nano-optoelectronics and biomedicine.

## 2. Methods
### 2.1 LR-TDDFT calculations
The optical and ground-state electronic properties of peropyrene and its derivatives have been conducted using LR-TDDFT. To achieve a deeper understanding of the plasmonic excitation characteristics of the system,[18] the Perdew-Burke-Ernzerhof (PBE) exchange-correlation functional,[50] the 6-311G(d,p) basis sets[51] and the DFT-D3(BJ) correction method[52,53] are used for the DFT calculations, including the geometric optimization, ground-state and absorption spectrum calculation. In addition, all the orbital transition contributions greater or equal to $10^{-5}$ to each excitation state are considered to ensure an accurate description of the excitation state. For charge-doped systems, the calculation method is applied as for neutral molecules. All LR-TDDFT calculations are done by Gaussian 16 software.[54] The ground-state electronic structure and the excited-state electronic properties are processed using the wavefunction analysis software Multiwfn.[55,56]

Additionally, two plasmonic properties analytical methods: PI and GPI are employed. Both methods are utilized to quantify the plasmonic characteristics of excitations in nanostructures. Specifically, in the case of PI,

$$PI = 10^{N^{0\to I}-1} \qquad (1)$$

$N^{0\to I}$ represents the total number of electrons that transition from the ground state (0) to the excited state (I):

$$N^{0\rightarrow I} = \sum_{i=N_{occ}+1}^{N_T} \sum_{a=1}^{N_{occ}} \left(X_{ai}^{I*}X_{ai}^{I} + Y_{ai}^{I}Y_{ai}^{I*}\right) \qquad (2)$$

where $N_T$ represents the total number of orbitals, $N_{occ}$ represents the number of occupied orbitals, $X_{ai}^{I}$ and $Y_{ai}^{I}$ represent the excitation and deexcitation coefficients, respectively. For a case of complete single-particle excitation,

$$N_{single}^{0\rightarrow I} = \sum_{i=N_{occ}+1}^{N_T} \sum_{a=1}^{N_{occ}} X_{ai}^{I*}X_{ai}^{I} = 1 \qquad (3)$$

which corresponds to PI = 1. When the contribution of the deexcitation process becomes apparent, the plasmonic nature of the system begins to emerge, with PI exceeding 1.

Subsequently, in the context of GPI,

$$\text{GPI} = \frac{E_{\text{plas},I}}{\Gamma} \qquad (4)$$

$$E_{\text{plas},I} = \iint dr_1\, dr_2 \frac{\rho_I(r_1)\rho_I(r_2)}{|r_1 - r_2|} \qquad (5)$$

where $\Gamma$ is the damping energy of the transition and $\rho_I(r_1)$ is the transition density for the excitation $I$. Within the TDDFT description, all electronic excitations share the same damping energy $\Gamma$, which is introduced as a constant broadening parameter. As a result, the GPI of a given resonance $I$ and its plasmonic energy $E_{\text{plas},I}$ are proportional to each other. In this work, we calculate and present the $E_{\text{plas},I}$ values instead of the GPI values, referring to them as GPI values for brevity.[57,58]

**2.2 RT-TDDFT calculations**

Furthermore, the optical properties and localized field enhancement under RT-TDDFT framework are considered, the DFT and TDDFT calculations are performed using the open-source GPAW code (version:22.8.0) package[59,60] based on the linear combination of atomic orbitals (LCAO) mode.[61] The PBE exchange-correlation functional in the adiabatic limit, the default projector augmented-wave (PAW) data sets[62] and double-$\zeta$ polarized (dzp) basis sets provided in GPAW are used for C, H, O and N. A grid spacing parameter of 0.3 Å is chosen to represent densities and potentials, and the molecules are surrounded by a vacuum region of at least 8 Å. Ensuring convergence in our calculations, we apply a Fermi-Dirac smearing of 0.05 eV to the occupation numbers. For the time propagation, we use the Crank-Nicholson propagator[63] with a time step of $\Delta t$ = 10 attoseconds (as) and total propagation time of at least T = 20 femtoseconds (fs). To account for broadening effects, Gaussian damping with a width parameter of 0.20 eV is applied to the spectra.

TCM is an extremely effective analytical tool for delving into the excitation characteristics of plasmonic excitations in the excited states. Studies have shown that plasmonic resonances are typically complex phenomenon resulting from the superposition of numerous electron-hole excitations. Additionally, the discrepancy between the electron-hole orbital energy level differences and the excited energy corresponding to the excited state reveals the extent of electron-electron interaction contributions in the excitation process. This discrepancy serves as a key indicator for assessing the nature of plasmonic excitation, providing important theoretical insights into understanding the excitation behavior of plasmons. Mathematically, the TCM is expressed as:

$$M_\omega^{\text{TCM}}(\varepsilon_o, \varepsilon_u) = \sum_{ia} w_{ia}(\omega) g_{ia}(\varepsilon_o, \varepsilon_u) \tag{6}$$

where $\varepsilon_o$ and $\varepsilon_u$ represent the energy regions for occupied and unoccupied states, and the discrete $i \to a$ transition contributions are broadened by a 2D Gaussian function

$$g_{ia}(\varepsilon_o, \varepsilon_u) = \frac{1}{2\pi\sigma^2} \exp\left[-\frac{(\varepsilon_o - \varepsilon_i)^2 + (\varepsilon_u - \varepsilon_a)^2}{2\sigma^2}\right] \tag{7}$$

Finally, TCM is also employed in the LR-TDDFT computational method.

## 3. Results and Discussion
### 3.1 Plasmonic Properties of Molecules in Excited States

In our preliminary study, we analyze the excitation properties of peropyrene, as well as its derivatives modified with nitrogen atoms (peropyrene-N) and those co-doped with nitrogen and oxygen atoms (peropyrene-O). By solving the time-dependent Kohn-Sham (TD-KS) equations, we can obtain the energies and wavefunctions of molecular excited states, which can then be used to compute properties such as absorption spectra. It is noteworthy that for all three molecular configurations, the excitation contributions are predominantly confined to the longitudinal (*x*-direction) and transverse (*y*-direction) orientations, while the contribution perpendicular to the molecular plane (*z*-direction) is almost negligible (see Figure S1 in the *Supporting Information* (SI) for more details). Consequently, the *z*-direction excitation contribution is not considered in the RT-TDDFT calculations.

Figure 1 displays the absorption spectra which exhibit a high consistent among them. Compared to the LR-TDDFT spectra, the RT-TDDFT spectra display a slight blue-shift. The inserts within Figure 1 depict the structural schematics of the molecules under investigation. Furthermore, we conducted a detailed examination of individual excited states derived from LR-TDDFT calculations, using two plasmonic properties analytical methods (PI and GPI) to investigate plasmonic properties. As illustrated in Figure 1, the analysis reveals the plasmonic characteristics of the two types are qualitatively in agreement. Here we pinpoint key excited states for in-depth analysis and discourse, with particular attention to those highlighted (short arrows) in Figure 1. Excited states marked by red circles are those with relatively high values, whereas those encircled in purple denote states with relatively low GPI values yet significant oscillator strengths. Detailed insights can be found in Table S1 in the SI. TCMs prove to be an invaluable tool for elucidating the plasmonic excitation properties of these states, therefore, we carefully process the LR-TDDFT calculation results and plot the TCMs for the key excited states.

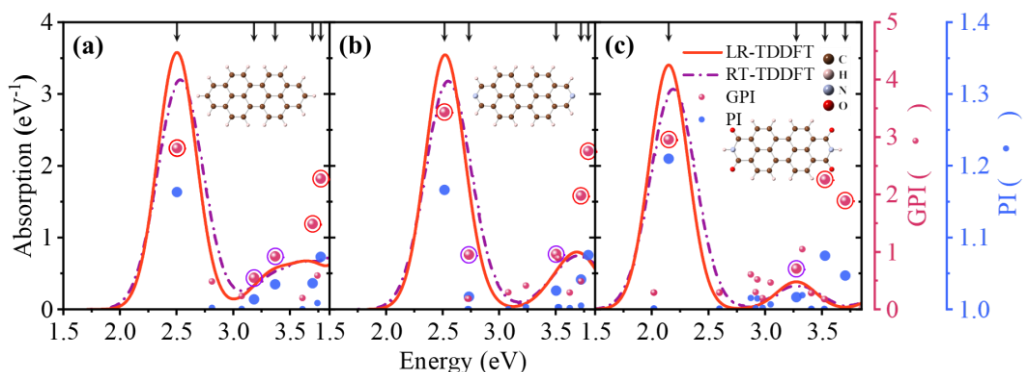

**Figure 1.** Absorption spectra and plasmonic properties of (a) peropyrene, (b) peropyrene-N, (c) peropyrene-O. The absorption spectra are obtained using LR-TDDFT and RT-TDDFT methods. The plasmonic properties are determined using PI and GPI methods. The inserts depict the corresponding molecular structural schematics. The excited states corresponding to the short arrows are the key excited states. Among them, red circles indicate excited states with relatively high GPI values, whereas purple circles denote excited states with relatively low GPI values but significant oscillator strengths.

Figure 2 shows the TCMs and DOS for the selected excited states of peropyrene, encompassing details on excitation energy and oscillator strength. Within the TCMs, positive and negative contributions reflect the phase alignment of orbital transition dipole moments relative to the overall transition dipole moment of the excited state, and the dipole moment components chosen for analysis are based on the predominant excitation direction of states under consideration. It can be noticed that the oscillator strength is proportional to the square of the transition dipole moment, where in-phase and out-of-phase only refer to a relative relationship. Moreover, the compact molecules exhibit distinct, well-separated KS states, a feature that is supported by the DOS diagrams.[37] Concurrently, the insets of Figure 2 also present the transition densities associated with the respective excited states.

Firstly, we investigate the $S_1$ ($S_0 \rightarrow S_1$, and so on), $S_4$, $S_5$, $S_7$ and $S_9$ excited states of peropyrene (the encircled states indicated by the short arrows in Figure 1). As depicted in Figure 2a, the primary orbital transition energy (the red halo) for the $S_1$ is notably lower than its excitation energy (the black dashed line), which points to a significant electron-electron interaction. This observation suggests that the $S_1$ is characterized by pronounced plasmonic excitation features.[18] The transition density reveals a dipole-symmetric distribution, aligning with the conventional notion of plasmons as collective excitations resembling standing density waves, thereby confirming that the $S_1$ corresponds to a longitudinal dipole plasmonic excitation. Furthermore, the DOS indicates that this excited state is predominantly involves a HOMO→LUMO orbital transition. Owing to its small molecular size, the KS states are discrete, resulting in less pronounced collective behavior in its plasmonic excitation. Additionally, Figure 2e indicates the $S_9$ also exhibits strong electron-electron interactions, despite having an oscillator strength of zero. The transition density suggests that it corresponds to a longitudinal standing wave dark plasmonic mode with two nodes. The plasmonic characteristics of these excited states can be further substantiated by their relatively higher PI and GPI values.

However, the $S_4$, $S_5$ and $S_7$ (see Figures 2b-2d) all exhibit contributions from multiple transition orbitals. Furthermore, the smaller transverse geometric dimension of the molecular suppresses the formation of plasmonic excitation modes, leading to intricate transverse excitation patterns mixed with other single-particle excitations. In the case of the $S_4$, the energy level difference of the main contributing transition orbital (the red halo in Figure 2b) is equivalent to the excitation energy (the black dashed line), whereas the differences for the other orbitals are marginally less than the excitation energy. This indicates very weak electron-electron interactions, suggesting that it is predominantly a single-particle excitation mode. Similarly, the $S_5$ corresponds to a mode that exhibits a strong mixture of single-particle and plasmonic excitations, while the $S_7$ is primarily attributed to a weak plasmonic excitation mode. This interpretation is supported by the observed trends in PI and GPI values. The presence of certain plasmonic excitation modes leads to the transition densities of the $S_4$, $S_5$ and $S_7$ all displaying transverse dipole excitation characteristics. Consequently, this observation underscores that the determination of whether an excited state corresponds to a plasmonic excitation mode cannot be based solely on the analysis of a single transition density. Instead, a comprehensive, multi-dimensional analysis from various perspectives is essential to fully elucidate the excitation characteristics of these excited states.

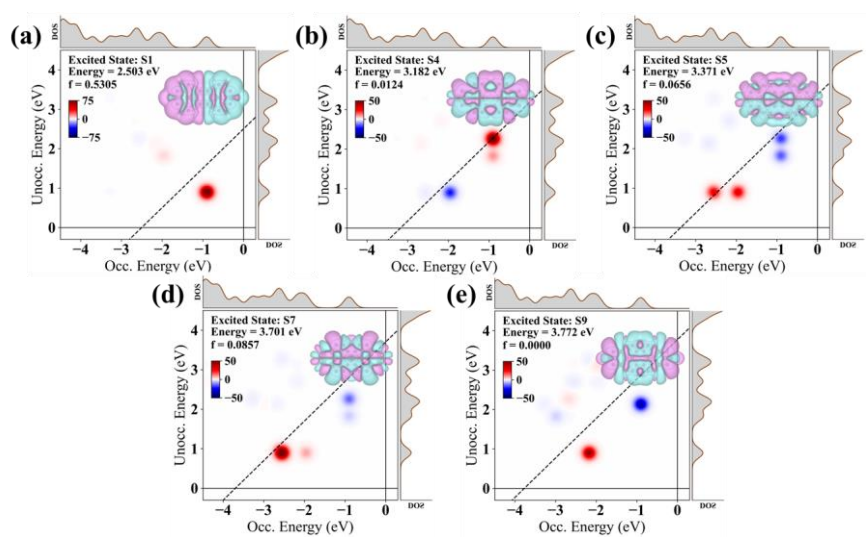

**Figure 2.** TCMs for key excitation states of peropyrene. (a) $S_1$, (b) $S_4$, (c) $S_5$, (d) $S_7$ and (e) $S_9$. The TCMs depict the orbital transition contributions to the excited states, with red signifying positive contributions and blue indicating negative contributions. The dashed line is included to represent a constant excitation energy level. The f values represent the oscillator strength. The insets show the transition density.

To investigate the molecular excitation properties under different elemental doping conditions, we extend our analysis beyond peropyrene to further examine the excited states of peropyrene-N and peropyrene-O, as presented in Figures S2 and S3 in the SI. The TCMs and transition densities clearly indicate that the $S_1$ and $S_{13}$ of peropyrene-N correspond to a longitudinal dipole plasmonic excitation mode and a longitudinal standing wave dark plasmonic mode with two nodes, respectively. Additionally, the $S_4$ exhibits strong characteristics of single-particle excitation, the $S_7$ is identified as a strong mixed excitation mode and the $S_{11}$ is primarily characterized by plasmonic excitation. Despite the introduction of nitrogen atoms alters the atomic composition of the molecular system, the overall conjugated structure remains largely intact. Consequently, the plasmonic excitation modes of peropyrene-N are largely preserved, suggesting a high degree of consistency to those of peropyrene. However, the excited states dominated by single-particle excitation exhibit notable changes due to the variations in atomic composition.

The further doping of oxygen atoms in peropyrene-O results in modifications to both the atomic composition and the conjugated structure, leading to substantial differences in both plasmonic and single-particle excitation when compared to peropyrene and peropyrene-N. In peropyrene-O, the $S_3$ and $S_{17}$ correspond to a longitudinal dipole plasmonic excitation and a longitudinal standing wave dark plasmonic mode with two nodes, respectively. The $S_{18}$ displays a more complex plasmonic excitation mode, akin to the quadrupole mode in the traditional plasmonic excitation,[64] while the $S_{11}$ is more indicative of a single-particle excitation mode. The characteristics of the excited states in peropyrene-N and peropyrene-O are further substantiated by the variation trends observed in PI and GPI values.

## 3.2 Analysis of Optical Properties

Within the molecular systems investigated, the longitudinal dipole plasmonic mode manifests as the dominant absorption peak, characterized by the most significant plasmonic excitation features, which correspond to the highest values of PI and GPI. In contrast, the other plasmonic excitation modes are classified as dark modes and do not exhibit optical activity. Consequently, our subsequent analytical efforts will be directed towards the examination of the longitudinal dipole plasmonic mode.

We present the absorption peak intensities and resonance energies of the longitudinal dipole plasmonic resonances for the three molecular systems, as depicted in Figures 3a and 3b. The electron localization function (ELF) calculated solely based on π molecular orbitals (ELF-π),[65] is a widely used real space function for revealing the delocalization of π electrons in molecules containing conjugated structures. Here, we compute the ELF-π using the LR-TDDFT results for the three molecules. To better visualize the ELF-π, we plot color-filled maps at 1.2 Å above the molecular plane, as depicted in Figures 3c-3e. By comparing the color scales within the figures, it is evident that the degree of π electron delocalization in peropyrene-N remains largely unchanged relative to peropyrene, suggesting that the conjugated structure of peropyrene-N is preserved. In contrast, the delocalization of the π region in peropyrene-O is significantly hindered at the positions indicated by the arrows in Figure 3e, indicating that the conjugated structure of peropyrene-O has been partially disrupted.

For peropyrene-N, a slight reduction in absorption peak intensity is noted when comparing the results to those of peropyrene (see Figure 3a). In stark contrast, the intensity for peropyrene-O demonstrates a marked decrease. This phenomenon can be attributed to the inherent nature of plasmonic resonance peak development, which is influenced by the number of conjugated electrons cumulative effects for collective plasmonic excitations. The doping of nitrogen atoms into the molecular system does not substantially alter the conjugated structure, thereby preserving an approximately constant number of valence electrons involved in the collective plasmonic oscillations. As a result, the absorption peak intensity remains largely unaltered. On the other hand, the further introduction of oxygen atoms disrupts the molecular conjugated structure, leading to a reduction in the valence electrons participating in the plasmonic excitations. This disruption consequently results in a significant decrease in the absorption peak intensity for peropyrene-O.[35]

The resonance energy of the plasmonic absorption peaks is notably influenced by the density of valence electrons. As shown in Figure 3b, while the doping of nitrogen atoms into the molecular system does not markedly modify the conjugated structure, it does induce a subtle contraction in the geometric dimensions. This contraction results in a minute enhancement of the valence electron density, and a slight blueshift in the plasmonic resonance energy is observed for peropyrene-N. In contrast, the additional doping with oxygen atoms leads to a smaller conjugated structure whose valence electrons mainly support the plasmon oscillation; and the electron density thus decreases, which in turn gives rise to a significant redshift in the resonance energy of peropyrene-O.

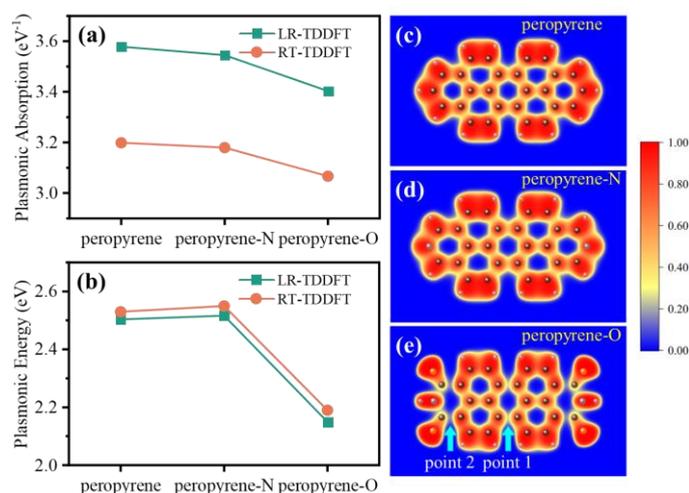

**Figure 3.** (a) Absorption peak intensities corresponding to the longitudinal dipole plasmonic resonance for peropyrene, peropyrene-N and peropyrene-O. (b) Resonance energies associated with the longitudinal dipole plasmonic resonance for peropyrene, peropyrene-N and peropyrene-O. (c-e) Color-filled maps of ELF-π at 1.2 Å above the molecular plane for (c) peropyrene, (d) peropyrene-N and (e) peropyrene-O.

The electromagnetic energy generated by the molecular plasmonic resonances can be confined to nanoscale regions, resulting in local enhancement of the external electromagnetic field near the nano systems, effectively functioning as optical plasmonic nano-antennas. Figures 4a-4c display field enhancement maps for peropyrene, peropyrene-N and peropyrene-O sliced along the $x = 0$, $y = 0$ and $z = 0$ directions at the molecular center, respectively. The maps reveal that peropyrene exhibits relatively considerable field enhancement effects within a range close to 3 Å in the $z$-direction. The incorporation of nitrogen atoms does not alter the plasmonic resonance characteristics of the system, hence the field enhancement properties remain largely unchanged, exhibiting high consistency with peropyrene. However, as observed in the analysis, the subsequent doping with oxygen atoms leads to modifications in the conjugated structure, which in turn results in significant alterations to the field enhancement characteristics. At the corresponding longitudinal dipole plasmonic resonance, peropyrene-O displays the maximum field enhancement at the edge of the molecule, while peropyrene and peropyrene-N exhibit the strongest field enhancement at the center of the molecular plane. Moreover, due to the reduced oscillating electron density, the maximum field enhancement for peropyrene-O is relatively weaker. This further indicates that changes in the conjugated structure lead to a transformation in the electron oscillation modes participating in the plasmonic resonance.

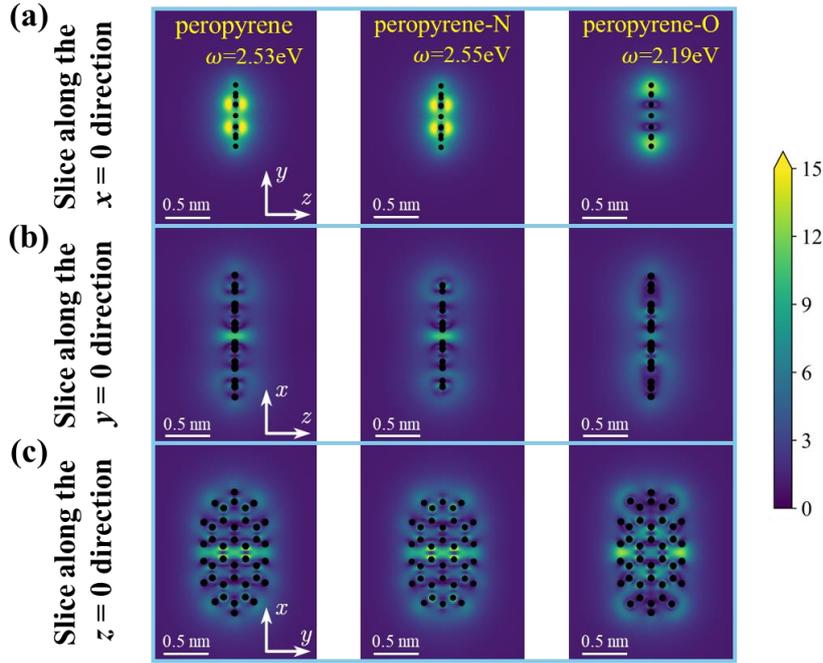

**Figure 4.** Electric field enhancement for peropyrene (left), peropyrene-N (middle) and peropyrene-O (right). The maps are sliced along the (a) $x = 0$, (b) $y = 0$ and (c) $z = 0$ directions at the molecular center, corresponding to the longitudinal dipole plasmonic resonance energy.

**3.3 Charge Doping Effects for Different Conjugated Structures**

We conduct a systematic analysis of the influence of electron and hole doping on the plasmonic excitations of the three molecules. Despite the emergence of new satellite peaks with weaker absorption intensities due to the charge doping, these peaks exhibit feeble plasmonic excitation characteristics. Consequently, our research focus is still directed towards the examination of longitudinal dipole plasmonic resonance modes. The absorption spectra, PI and GPI values under the charge doping conditions are shown in Figure 5, with corresponding parameters provided in Table S2 in the SI. As electron and hole doping proceeds, alterations in the occupied and unoccupied molecular orbitals occur, resulting in a shift of the dipole plasmonic resonance peaks. A comparative analysis reveals that charge doping exerts a pronounced effect on the plasmonic resonance of peropyrene and peropyrene-N, with the overall resonance peaks exhibiting a red-shift trend as the level of charge doping increases. Moreover, the impact of electron doping is observed to be more substantial than that of hole doping. In contrast, peropyrene-O does not display a consistent trend, instead, it undergoes a red-shift followed by a blue-shift with increasing charge doping. This behavior further suggests that the incorporation of oxygen atoms has a more marked influence on the molecular orbital structure. Previous works have indicated that electron and hole doping can yield distinct outcomes in graphene flakes due to the asymmetry of their energy bands,[49,66,67] the divergent effects of electron and hole doping on molecules can be attributed to the unique manner in which each type of doping modifies the intrinsic molecular characteristics.

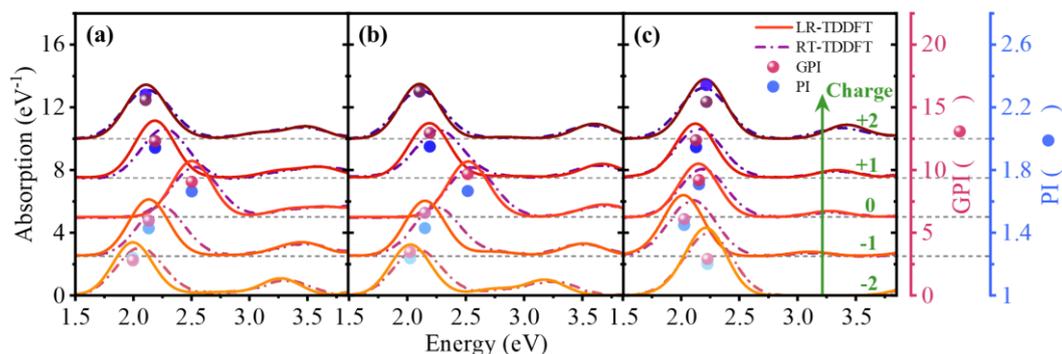

**Figure 5.** Absorption spectra and plasmonic properties of (a) peropyrene, (b) peropyrene-N and (c) peropyrene-O under charge doping. The spectra, from bottom to top, correspond to doping levels ranging from two additional electrons to two additional holes.

The ELF-π values at distinct bifurcation points, indicated by the arrows in Figure 3e, under various charge doping conditions are provided in Table 1. These bifurcation points are identified as the junctures where the continuity of an ELF-π isosurface is disrupted, leading to the formation of two separate entities upon elevation of the isovalue. It is believed that a higher ELF-π value at a bifurcation point signals a higher degree of facilitation for π electron delocalization between the contiguous ELF domains.[68] It can be observed that with the increase in charge doping levels, the degree of π electron delocalization in peropyrene and peropyrene-N is hindered to some extent, leading to the disruption of the conjugated structures, which in turn results in a redshift of the plasmonic resonance energy. However, peropyrene-O exhibits a different trend. Overall, as the level of charge doping increases for peropyrene-O, the degree of π electron delocalization is enhanced to a certain degree. The slight blueshift of the resonance energy under ±2 charge doping conditions, compared to the neutral molecule, confirms this phenomenon.

**Table 1.** The ELF-π values at distinct bifurcation points for peropyrene, peropyrene-N and peropyrene-O under charge doping. These points are valuable in quantifying the strength of π conjugation of molecules.

| peropyrene | | | peropyrene-N | | | peropyrene-O | | |
|---|---|---|---|---|---|---|---|---|
| charge | point 1 | point 2 | charge | point 1 | point 2 | charge | point 1 | point 2 |
| +2 | 0.472 | 0.762 | +2 | 0.477 | 0.775 | +2 | 0.681 | 0.319 |
| +1 | 0.605 | 0.795 | +1 | 0.608 | 0.806 | +1 | 0.576 | 0.291 |
| 0 | 0.723 | 0.819 | 0 | 0.727 | 0.831 | 0 | 0.461 | 0.338 |
| -1 | 0.569 | 0.742 | -1 | 0.569 | 0.755 | -1 | 0.592 | 0.464 |
| -2 | 0.446 | 0.660 | -2 | 0.445 | 0.676 | -2 | 0.690 | 0.580 |

In the context of charge doping, the dimensions of these molecules remain largely invariant. Furthermore, PI value is inherently a dimensionless parameter. Consequently, we can utilize PI values to conduct a more detailed analysis and comparison of the plasmonic excitation properties in molecular systems under charge doping conditions. As depicted in Figure 5 and detailed in Table S2 in the SI, PI values for peropyrene and peropyrene-N exhibit a progressive increase in correlation with the extent of charge doping, with hole doping exerting a more pronounced effect than electron doping. This observation contrasts with the previously noted trend of spectral redshift associated with charge doping. However, in the case of peropyrene-O, PI values remain relatively stable, apart from a notable change upon the doping of two holes. The spatial distribution of the induced charge density is a well-recognized indicator of plasmonic oscillations. To this end, we conduct an integration of the induced charge density along the *x*-direction and across the entire space for three molecules under various charge doping conditions at the respective longitudinal dipole plasmonic resonance energy, as shown in Figure 6. The results suggest that for peropyrene and peropyrene-N, the induced charge density progressively spreads toward the molecular edges with electron and hole doping. This phenomenon indicates intensified plasmonic oscillations, which in turn lead to a rise in PI values. This behavior is not significantly observed in peropyrene-O, accounting for the minimal variation in its PI values with charge doping.

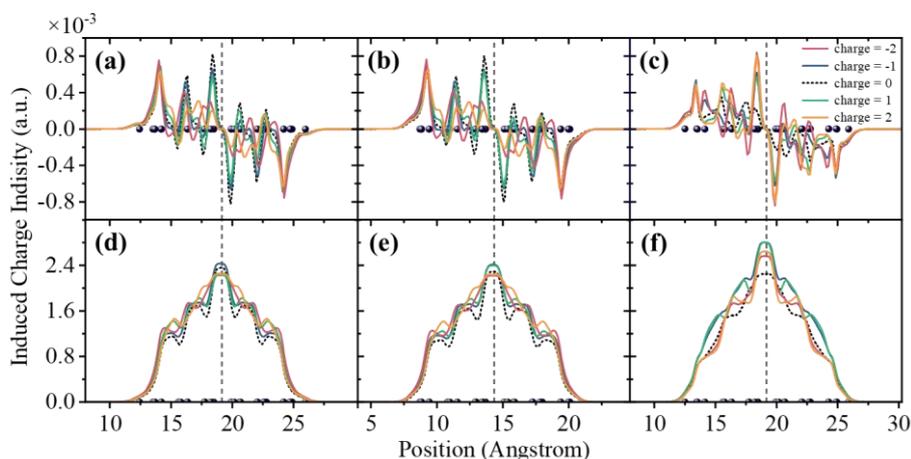

**Figure 6.** Induced charge density integral curves for three molecules under charge doping. (a-c) show the integrals along the *x*-direction for peropyrene, peropyrene-N and peropyrene-O, respectively. (d-f) depict the comprehensive integrals across the entire space for peropyrene, peropyrene-N and peropyrene-O, respectively. All integrals are calculated at the respective longitudinal dipole plasmonic resonance energy.

Furthermore, the TCMs for the molecular longitudinal plasmonic resonances under charge doping are generated utilizing both the LR-TDDFT and the RT-TDDFT methods. These mappings are employed for comparative study to ascertain the robustness of the findings presented in the preceding sections. The TCMs for the three molecules are depicted in Figures S4-S9 in the SI (RT-TDDFT results for comparison in Figures S5, S7 and S9 in the SI). The results show that the transition densities and induced charge densities demonstrate an elevated level of agreement, suggesting that the plasmonic resonances derived from both methods share identical excitation characteristics and correspond to the same dipole plasmonic resonances. The molecular orbitals that predominantly contribute to the TCMs, as well as the distribution of their contribution strengths, are highly coherent, indicative of robust electron-electron interactions. This reinforces the reliability of the TCMs analysis results discussed previously, which are carried out using the LR-TDDFT method. Despite there are discrepancies in the methodologies for determining the positive and negative contributions between the two computations, these differences do not compromise the validity of our analysis here.

The tunability of molecular plasmonic excitation and energy shifts in peropyrene and its derivatives offers potential applications in the field of electrochromics. In an experimental setting, these molecules can be dissolved into electrically conductive transparent polymer gels to fabricate molecular plasmonic electrochromic devices (ECDs). Additionally, we have computed the color changes associated with the absorption spectra for the three molecules, as depicted in Figure S10 in the SI. Peropyrene and peropyrene-N demonstrate significant electrochromic tunability, whereas peropyrene-O exhibits robust color stability, thereby expanding the range of achievable color tunability in experimental settings. It worth mentioning that the absorption spectra are influenced by the computational methods employed, and in practical experiments, they may also be affected by the specific choice of conductive transparent polymer gels.[38] Consequently, the actual colors observed in experimental settings may vary to some extent, however, the general trend of color tuning as a function of charge state is maintained.

## 4. Conclusions and Outlook

In this study, we investigate the plasmonic resonance characteristics of peropyrene and its derivatives influenced by the conjugated structures. By utilizing PI and GPI values in conjunction with TCMs, we elucidate the discrimination in mixed excited states and plasmonic excitation modes. The intricate nature of the excited state characteristics at the small molecular scale complicates the use of individual transition densities or induced charge densities as direct criteria for assessing plasmonic properties. Doping with nitrogen atoms does not markedly alter the number and density of conjugated electrons participating in dipole oscillations, resulting in a slight blue-shift in the dipole plasmonic resonance energy. In contrast, oxygen atoms doping significantly disrupts the conjugated structure, reducing the number and density of electrons, which in turn leads to a more substantial decrease in intensity and a red shift in resonance energy. Furthermore, the electromagnetic field induced by molecular plasmonic resonance is confined to the nanoscale region, yielding enhanced local electric field properties and acting as a molecular nano-antenna. At the corresponding longitudinal dipole plasmonic resonance, peropyrene-O displays the maximum field enhancement at the edge of the molecule, while peropyrene and peropyrene-N exhibit the strongest field enhancement at the center of the molecular plane. Charge doping induces a certain degree of alteration in the conjugated structures. The longitudinal dipole resonance peaks of peropyrene and peropyrene-N exhibit a gradual red-shift with increasing charge doping levels due to the π electron delocalization is hindered, indicating a broad tuning range, whereas the impact of charge doping on peropyrene-O is relatively weak.

These findings highlight the high sensitivity of molecular plasmonic properties to conjugated structures and demonstrate the significant utility of multi-method analytical strategy in explaining such complex phenomena. The alterations in plasmonic properties induced by the incorporation of oxygen atoms clearly illustrate the potential tunability of peropyrene-based materials in the field of molecular plasmonic technology. Moreover, the analytical strategy employed in this study are universally applicable, extending beyond the scope of the current research. Through this analytical approach, we can provide a thorough understanding of the plasmonic excitation properties in molecules and other nano systems, which is of vital importance for advancing their applications in the field of plasmonic and related technologies.

**Authors contributions**
Y. F conceived the idea and directed the project. H. L and N. G did DFT and TDDFT calculations. H. L and Y. F analyzed the data and wrote the manuscript. All the authors revised the manuscript.

**Funding**
This research was supported by the National Natural Science Foundation of China (Grant No. 12274054, 12074054).

**Conflicts of interest**
The authors declare no competing financial interest.

**Availability of data and material**
The data and material that support the findings of this study are available from the corresponding author upon reasonable request.